# MHC Restriction of V-V Interactions in Serum IgG

Earnest Leung[1] and Geoffrey W. Hoffmann[1*]

**According to Jerne's idiotypic network hypothesis, the adaptive immune system is regulated by interactions between the variable regions of antibodies, B cells, and T cells.[1] The symmetrical immune network theory[2,3] is based on Jerne's hypothesis, and provides a basis for understanding many of the phenomena of adaptive immunity. The theory includes the postulate that the repertoire of serum IgG molecules is regulated by T cells, with the result that IgG molecules express V region determinants that mimic V region determinants present on suppressor T cells. In this paper we describe rapid binding between purified murine serum IgG of a given strain and IgG of the same strain and also between IgG from MHC-matched mice, but not between serum IgG preparations of mice with different MHC genes. We interpret this surprising finding in terms of a model in which IgG molecules are selected to have both anti-anti-(self MHC class II) and anti-anti-anti-(self MHC class II) specificity.**

## Theory

The symmetrical immune network theory of the regulation of the adaptive immune system[2,3] is based on Jerne's immune network hypothesis.[1] The theory resolves several paradoxes and makes several experimentally testable predictions. The paradoxes are phenomena that do not make any sense in the context of clonal selection without taking idiotypic network interactions into account. These include for example low dose tolerance including ultra-low dose tolerance, the existence of antigen-specific suppressor T cells, antigenic competition, the IJ paradox and the Oudin-Cazenave paradox. One of the predictions is that anti-IJ$^A$ antibodies produced in an animal of strain B immunized with A strain lymphocytes are specific for anti-anti-A antibodies produced in an strain A animal immunized with B strain lymphocytes and vice versa.

As part of the theory we have previously proposed IgG is a quasi-species, meaning that the IgG molecules of an unimmunized vertebrate have V regions that are similar to each other.[2] The theory specifies that V regions of IgG antibodies mimic the V regions on CD8 suppressor T cells. In this paper we report MHC restriction of V-V interactions in serum IgG of mice, which we interpret in terms of the IgG being selected to have complementarity to both anti-(MHC class II) helper T cells and anti-anti-(MHC class II) suppressor T cells. In other words, normal IgG is both anti-anti-(MHC class II) and anti-anti-anti-(MHC class II). This interpretation defines normal IgG more sharply and in this context is new evidence supportive of IgG being a quasi-species.

[1]Network Immunology Inc., 3311 Quesnel Drive, Vancouver, B.C. Canada V6S 1Z7
*Correspondence to hoffmann@networkimmunologyinc.com

A recurring theme in the symmetrical network theory is co-selection.[4] For example, suppressor T cells are a central regulating element of the system, and they are co-selected with helper T cells. Co-selection means that the suppressor T cells are selected on the basis of having V regions with complementarity to the V regions of as many helper T cells as possible, and conversely helper T cells are selected such that their V regions have complementarity to the V regions of as many suppressor T cells as possible. An additional and well established constraint on the selection of Th1 helper T cells is that they are selected such that their V regions have complementarity to self MHC class II molecules.

Figure 1 is a model for the interactions between (a) MHC class II, (b) Th1 and Ts1 cells that are anti-(MHC class II), (c) Th2, Ts2 and IgG that are anti-anti-(MHC class II), and (d) Ts3 cells that are anti-anti-anti-(MHC class II), as published in reference 2 as Figure 17-2. In this model normal IgG is anti-anti-(self MHC class II). Ts1 cells may be Treg cells. Ts2 cells are classical CD8 suppressor T cells. It was suggested that this co-selection model could result in normal serum IgG being a quasi-species, with anti-anti-(self MHC class II) V regions that are similar to anti-anti-(self MHC class II) epitopes present on Ts2 suppressor T cells. In this paper we present evidence consistent with a model in which IgG V regions are more narrowly defined than that, with complementarity to both Th1 V regions and Ts2 V regions. In other words, IgG is selected to be both anti-anti-(self MHC class II) and anti-anti-anti-(self MHC class II). This dual specificity of serum IgG results in IgG from a given strain of mice binding rapidly to serum IgG from the same strain of mice and to IgG from mice with the same MHC haplotype.

**Experiment**

**Methods Summary**

**Animal Care:** During this study, the care, housing and use of animals was performed at the Zoology Small Mammal Unit, Department of Zoology, University of British Columbia (UBC), in accordance with the Canadian Council on Animal

Care guidelines. The methodology described here was reviewed and approved by the UBC Committee on Animal Care prior to conducting the studies.

**IgG purification:** BALB/cJ, C57BL/6J, B10, and B10.D2 mice were sacrificed by $CO_2$ asphyxiation and whole blood collected by cardiac puncture. Serum was collected by centrifugation at 3000g and IgG collected with the Melon™ Gel IgG Spin Purification Kit (Pierce, Rockford, Il). Purity was confirmed by Western Blot, and IgG concentration determined by Coomassie Blue staining.

**Biotinylated IgG Preparation:** Purified IgG was diluted in oxidation buffer (0.1 M sodium acetate buffer, pH 5.5) to 2 mg/mL. One to one cold sodium meta-periodate solution to cold IgG was mixed and the reaction vessel was protected from light and incubated for 30 minutes at 4°C. Excess periodate was removed by gel filtration through a Zebra™ Desalt Spin Column equilibrated with coupling buffer (Pierce.) One to nine parts prepared 50 mM Biotin Hydrazide Solution to 9 parts oxidized and buffer-exchanged sample (results in 5 mM Biotin Hydrazide) was mixed for 2 hours at room temperature. Biotinylated IgG was separated from non-reacted material by Zeba™ Desalt Spin Columns. Biotinylated samples were aliquoted and stored at -70C. The concentrations of the stock solutions were 12.5 mg/mL for BALB/c biotin-IgG and 27 mg/mL for C57BL/6 biotin-IgG. For the experiment of Figure 1 the concentrations of biotin-IgG used were the stock solutions diluted by a factor of $10^8$.

**HRP-conjugated IgG Preparation:** Purified IgG was conjugated at pH 7.2. Purified IgG was diluted in PBS to 2 mg/mL. 10 mg/mL EZ-Link® Plus Activated Peroxidase with 100 μl of ultrapure water was added to the IgG solution. 10 μl of Sodium Cyanoborohydride was immediately added to the reaction and incubated for 1 hour in a fume hood at room temperature. 20 μl of Quenching Buffer was added and reacted at room temperature for 15 minutes. Conjugated samples were aliquoted and stored at -70C. The concentrations of the stock solutions were 19.1 mg/mL for BALB/c HRP-IgG and 29 mg/mL for C57BL/6 HRP-IgG. For the experiment of Figure 1 the concentrations of HRP-IgG used were the stock solutions diluted by factors of $10^6$, $10^7$, $10^8$, $10^9$ and $10^{10}$. For the experiments of Figure 2 the concentrations of HRP-IgG used were the stock solutions diluted by factors of $10^7$, $10^8$ and $10^9$.

**ELISA Assays:** Binding of the IgGs to each other was measured in ELISA assays using avidin-coated plates. Various concentrations of biotin-coupled IgG were incubated on the plates at room temperature for 30 minutes, and excess reagent washed off with PBS for 3 x 5 min. HRP-coupled IgG was incubated on the plates for 30 min, 1, 3 or 18 h and unbound IgG washed off with PBS for 3 x 5 min. 1-Step™ Turbo TMB-ELISA reagent was added and incubated at room temperature for 30 min and stop solution added. Absorbance was read at 450 nm. The results are presented as mean and standard deviation for eight-fold replicas.

**Results**

IgG was purified from the serum of BALB/cJ, C57BL/6J, B10, and B10.D2 mice. Biotin and horse radish peroxidase (HRP) were coupled to aliquots of the purified IgG to produce BALB/cJ biotin-IgG, C57BL/6J biotin-IgG, B10 biotin-IgG, B10.D2 biotin-IgG, BALB/cJ HRP-IgG, C57BL/6J HRP-IgG, B10 HRP-IgG, and B10.D2 HRP-IgG.

Binding of the IgGs to each other was measured in ELISA assays using avidin-coated plates. Various concentrations of biotin-coupled IgG were incubated on the plates and unbound IgG was washed off with PBS. HRP-coupled IgG was incubated on the plates for 30 min, 1, 3 or 18 h and unbound IgG was washed off with PBS. An HRP substrate was added, the plates were incubated at room temperature, then the changes in optical density were determined. The results shown in the figures are the mean and standard deviation for eight-fold replicas.

Results for binding times of 30 minutes, 1 hour, 3 hours and 18 hours for the binding of BALB/c IgG on BALB/c IgG, BALB/c on C57BL/6, C57BL/6 on BALB/c and C57BL/6 on C57BL/6 are shown in Figure 2. The results for 30 minutes and 1 hour show that there is rapid binding of BALB/c IgG on BALB/c IgG and C57BL/6 IgG on C57BL/6 IgG, and no rapid binding of BALB/c IgG on C57BL/6 IgG or vice versa (30 minute and 1 hour time points). On the other hand, at 3 hours and 18 hours binding of BALB/c IgG on C57BL/6 IgG and vice versa emerges.

Additional results for 1 hour binding, that include the MHC congenic strains B10 ($H-2^b$) and B10.D2 ($H-2^d$), are shown in Figure 3. IgG from BALB/c ($H-2^d$) binds rapidly to IgG from B10.D2, but not to IgG from B10, and conversely IgG from C57BL/6 ($H-2^b$) binds rapidly to B10 but not to B10.D2. Furthermore, IgG from B10 binds rapidly to IgG from B10 but not to IgG from B10.D2, and IgG from B10.D2 binds rapidly to IgG from B10.D2 but not to IgG from B10. These results are evidence that the rapid self-binding phenomenon seen at the 1 hour time point is restricted by genes in the major histocompatability complex.

**Interpretation**

The major histocompatability complex is known to play a central role in the repertoires of T cells,[5] but has not previously been shown to impact on the repertoires of antibodies. In order to interpret these results we begin by reviewing some of the basic features of the symmetrical immune network theory. Specific T cell factors[6,7] play a central role in the theory.[2] For the sake of brevity we call them tabs. Tabs have a molecular weight of about 50,000, and are able to exert potent regulatory effects on the adaptive immune system.[6] There is evidence that tabs are cytophilic for non-specific accessory cells (A cells) including macrophages.[7] When bound to A cell surfaces, antigen-specific tabs are assumed to be able to stimulate antiidiotypic T and B cells, and antiidiotypic T cells are assumed to be able to stimulate antigen-specific T cells and B cells. A mixture of antigen-specific and antiidiotypic T cells on A cell surfaces would be able to stimulate T and B cells that are both antigen-specific and antiidiotypic. Th1 lymphocytes are selected to have some complementarity to MHC class II molecules.

The MHC restriction seen in our results is most simply understood in terms of the model of Figure 4. MHC class II antigens stimulate Th1 cells. These cells secrete tabs that have some affinity for MHC class II, and bind to the surfaces of A cells, where they stimulate Ts2 suppressor T cells. The latter secrete anti-anti-(MHC class II) tabs, resulting in a mixture of anti-(MHC class II) and anti-anti-(MHC class II) tabs being present on A cell surfaces. Hence there is co-selection of the Th1 cells and Ts2 cells that is catalyzed by the tabs on A cells. This leads to the selection of IgG producing B cells that are both anti-anti-(MHC class II) and anti-anti-anti-(MHC class II). These IgG molecules are then able to exhibit MHC restricted binding to each other. This is a very rapid process, which is interpreted as being due to all of the IgG antibodies being selected to have this dual specificity, so that the antibodies quickly find a complementary antibody, in spite of the low concentrations of the antibodies used in the assays.

We ascribe the slower binding between BALB/c IgG and C57BL/6 IgG, that is seen at 3 hours and 18 hours, to the great diversity of BALB/c V regions and the corresponding diversity of C57BL/6 V regions, with antibodies needing the longer time to diffuse and encounter by chance antibodies with complementary specificity, independent of any self-specific MHC-restricted interactions. The fact that the MHC restricted interaction at short times before the non-specific (not MHC restricted) binding is seen, is consistent with the interpretation that all of the serum IgG antibodies are both anti-anti-(self MHC class II) and anti-anti-anti-(self MHC class II). This interpretation leads to the prediction that the MHC restriction we see can be mapped to the MHC class II region of the MHC.

Our experiments include criss-cross specificity controls, and in this regard the results are very clean. While we detected the fast binding effect down to extremely low concentrations, the experiment did not go to concentrations that were low enough for the effect to titrate out. It is however logically impossible that the effect would not titrate out at sufficiently high dilutions.

The finding that IgG exhibits MHC-restricted self-binding in H-2$^b$ and H-2$^d$ mice is compelling evidence of the importance of idiotypic network regulation of the adaptive immune system. We have shown here that it can be understood in the context of the symmetrical immune network theory.[2,3] The data is consistent with normal IgG being a more narrowly defined quasi-species than previously expected, namely all of the IgG antibodies in an unimmunized animal being selected to be both anti-anti-(self MHC class II) and anti-anti-anti-MHC class II.

**Author Information**

The authors declare competing financial interest. Correspondence should be addressed to the Geoffrey W. Hoffmann (hoffmann@networkimmunologyinc.com).

**Figure captions**

Figure 1. An idiotypic network model of immune system regulation that includes MHC class II, Th1 and Th2 helper T cells, Ts1, Ts2 and Ts3 suppressor T cells and serum IgG antibodies. From reference 2, chapter 17.

Figure 2. ELISA assay results for the binding of binding of BALB/c serum IgG on BALB/c serum IgG, BALB/c on C57BL/6, C57BL/6 on BALB/c, and C57BL/6 on C57BL/6 at a) 30 minutes b) 1 hour c) 3 hours and d) 18 hours.

Figure 3. ELISA assay results for the binding at 1 hour of serum IgG to serum IgG involving purified IgG from B10, BALB/c, C57BL/6 and B10.D2 mice.

Figure 4. A co-selection model that explains the phenomenon of MHC restriction of V-V interactions in serum IgG. α is an abbreviation for "anti-". MHC class II antigens stimulate Th1 cells. These cells secrete tabs that have some affinity for MHC class II, and bind to the surfaces of A cells, where they stimulate Ts2 suppressor T cells. The latter secrete anti-anti-(MHC class II) tabs, resulting in a mixture of anti-(MHC class II) and anti-anti-(MHC class II) tabs being present on A cell surfaces. There is co-selection of the Th1 cells and Ts2 cells that is catalyzed by the tabs on A cells. This leads to the selection of IgG producing B cells that are both anti-anti-(MHC class II) and anti-anti-anti-(MHC class II). The IgG molecules then exhibit MHC restricted binding to each other.

Figure 1

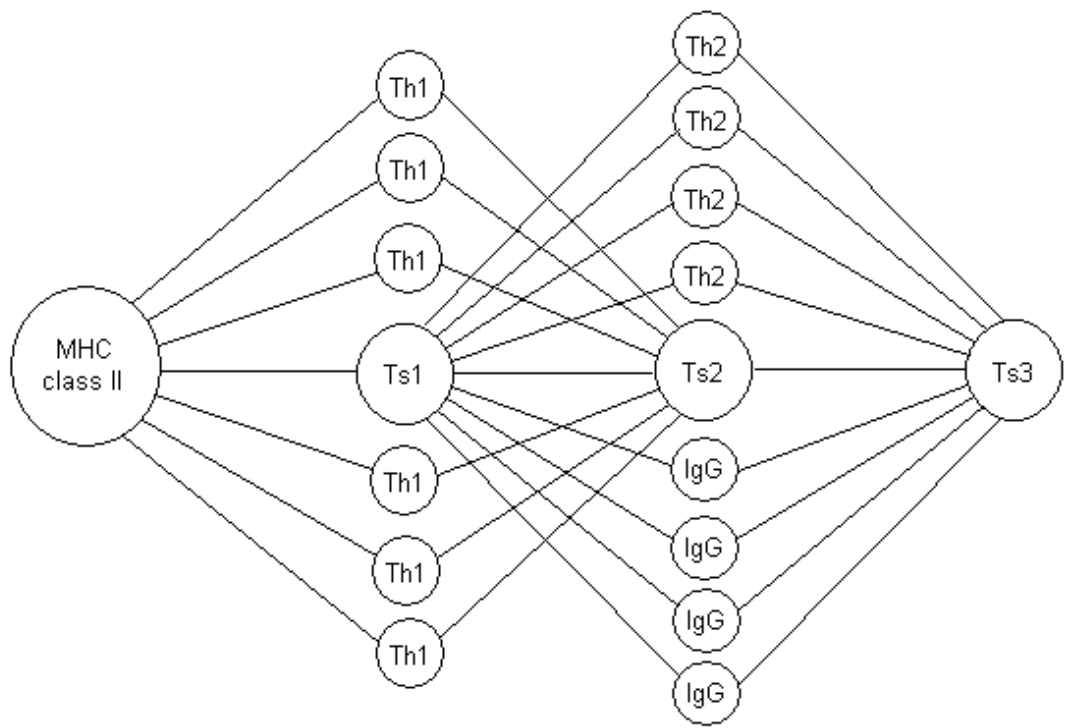

Figure 2 a)

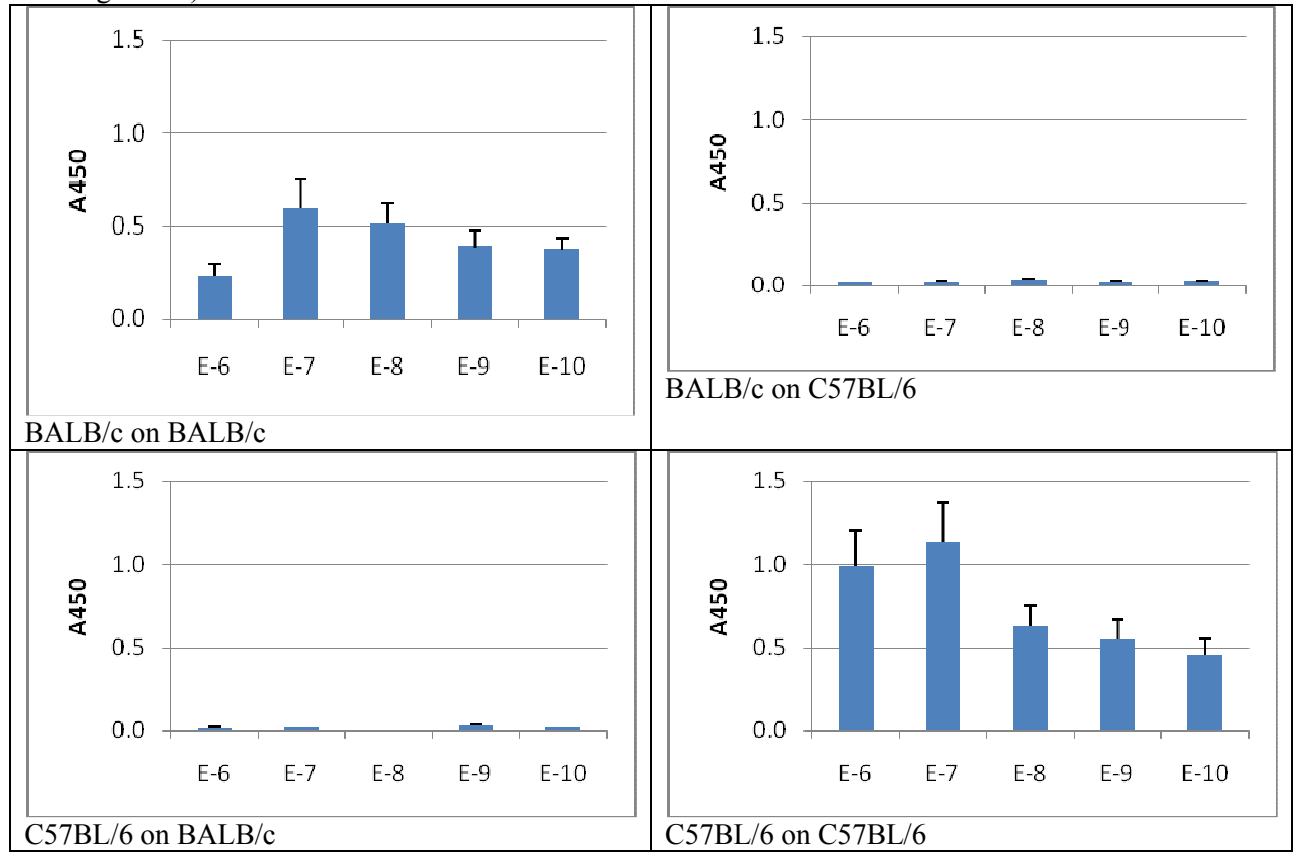

BALB/c on BALB/c

BALB/c on C57BL/6

C57BL/6 on BALB/c

C57BL/6 on C57BL/6

Figure 2 b)

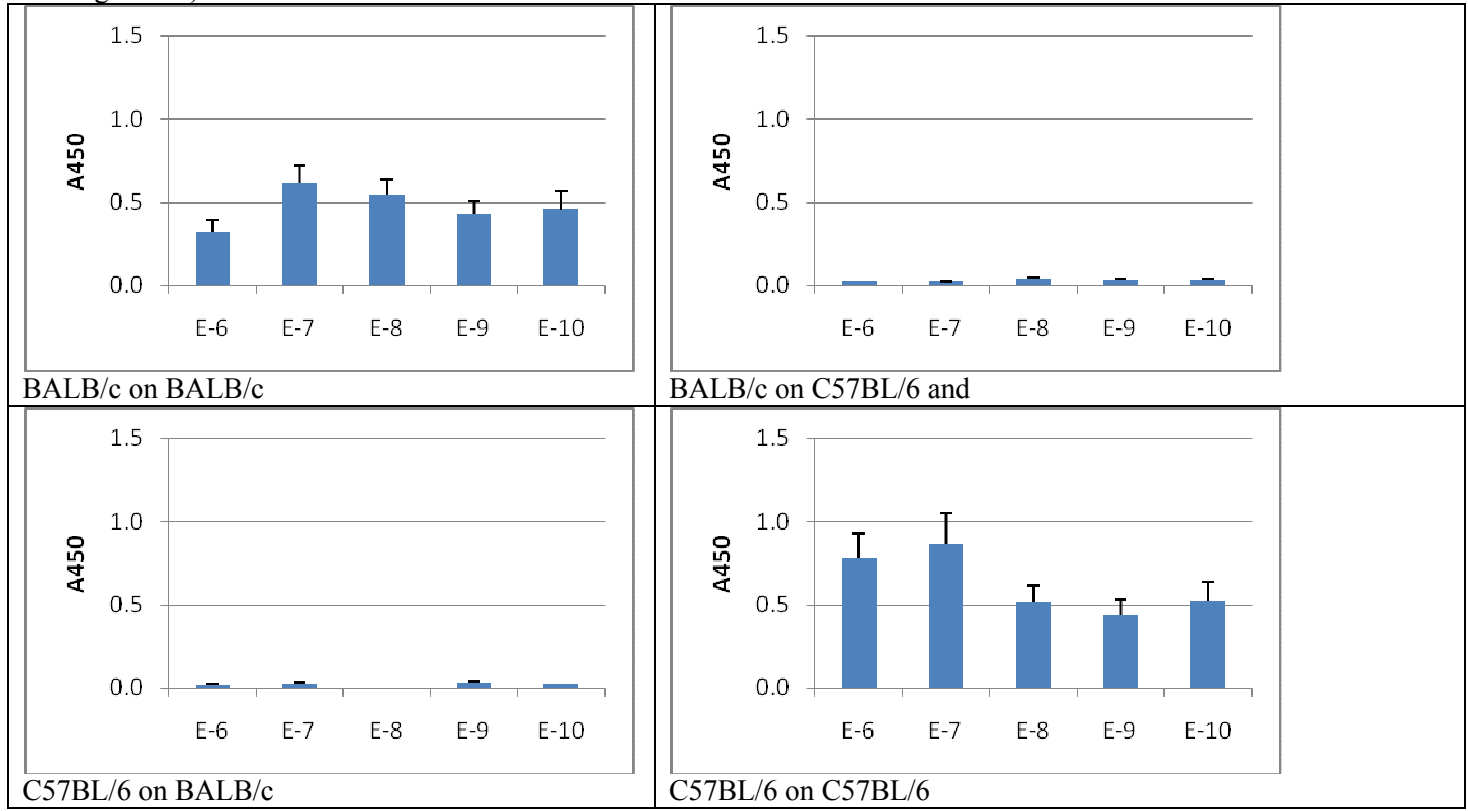

Figure 2 c)

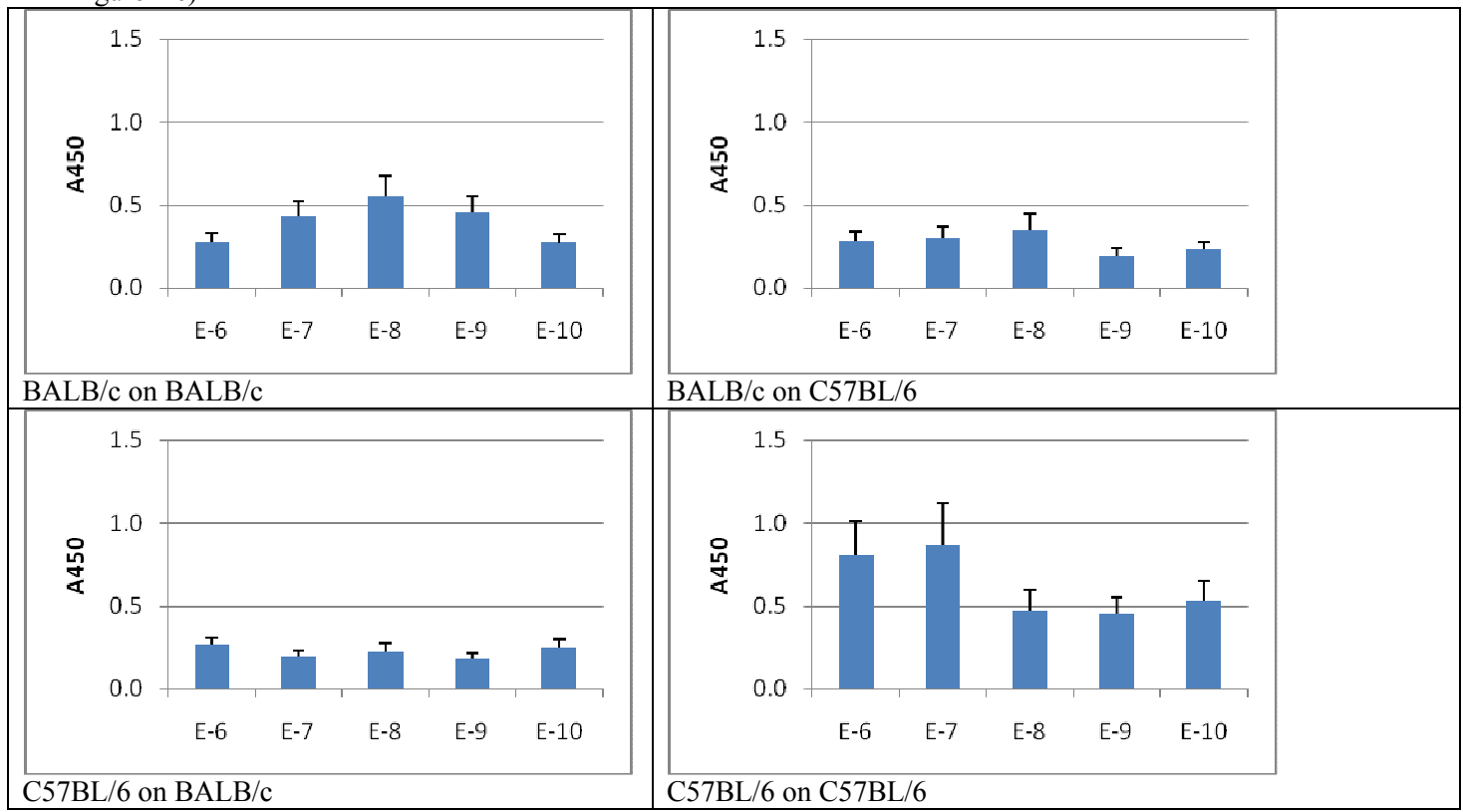

BALB/c on BALB/c

BALB/c on C57BL/6

C57BL/6 on BALB/c

C57BL/6 on C57BL/6

Figure 2 d)

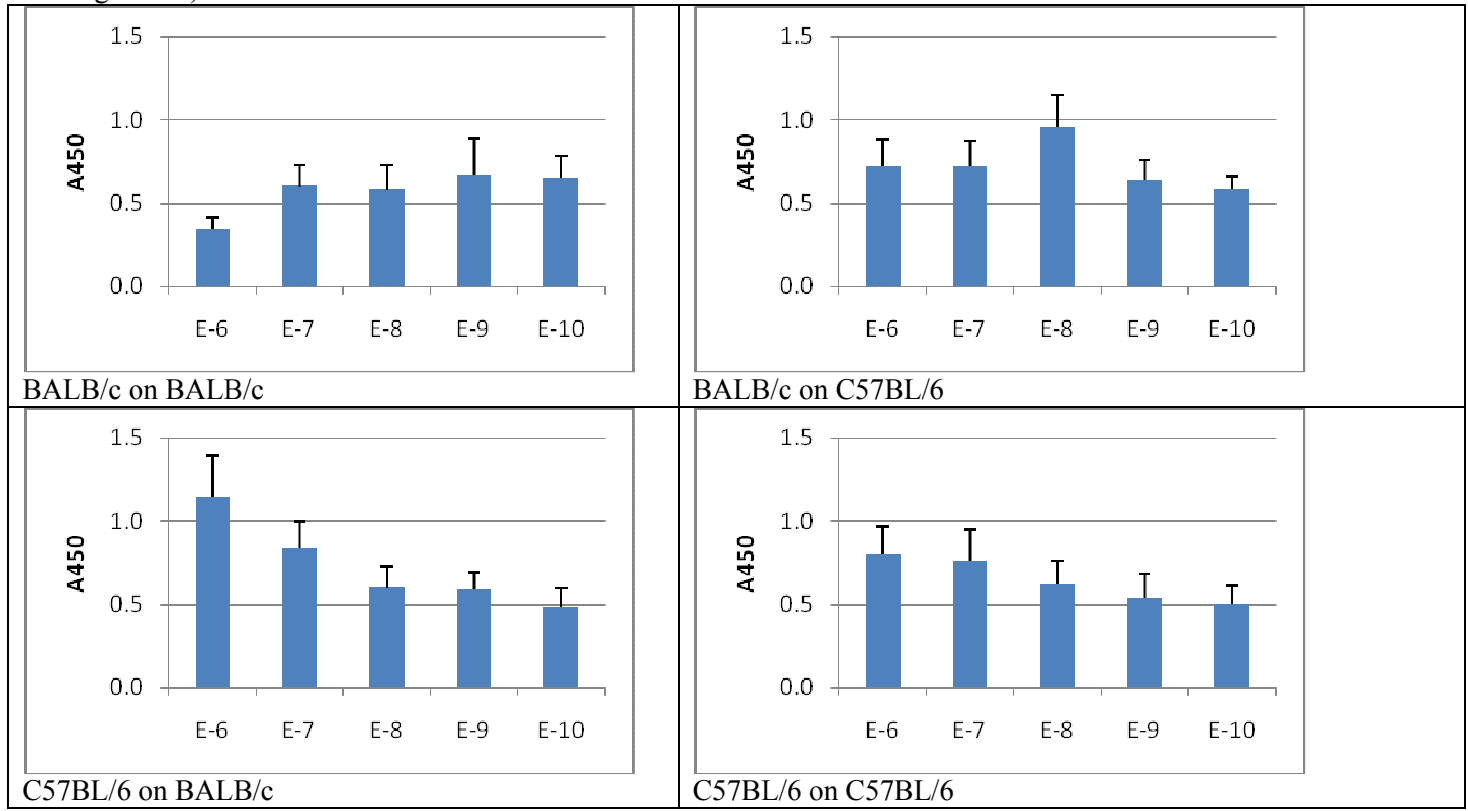

Figure 3

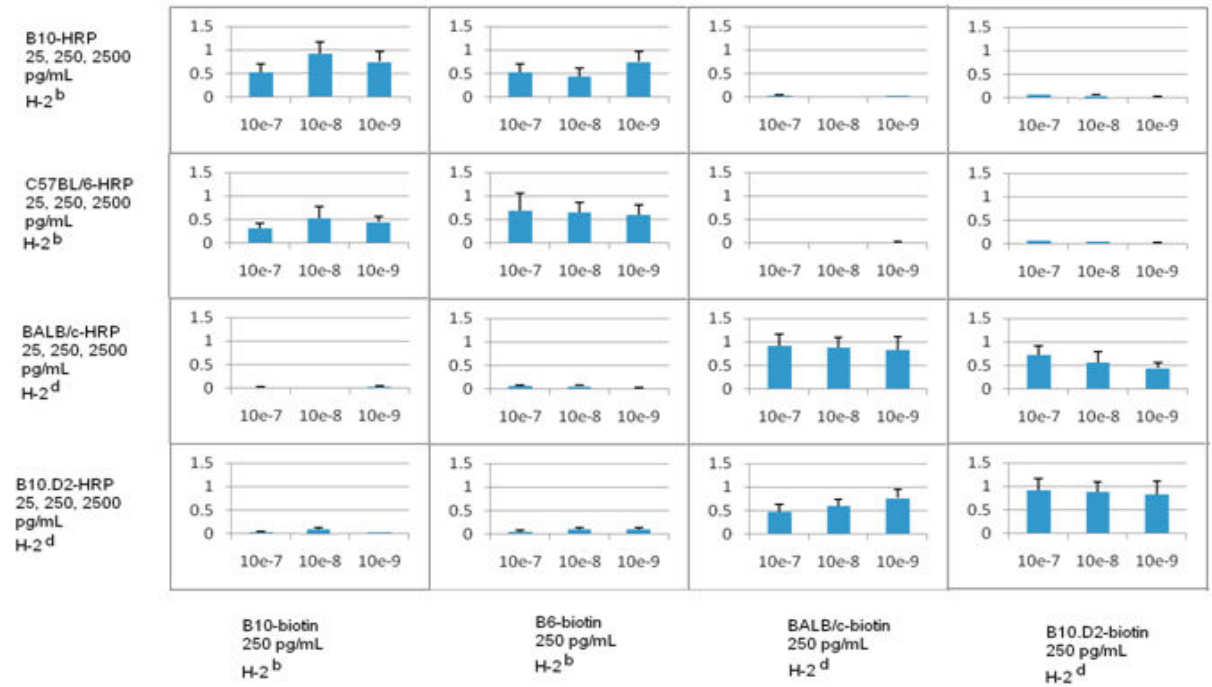

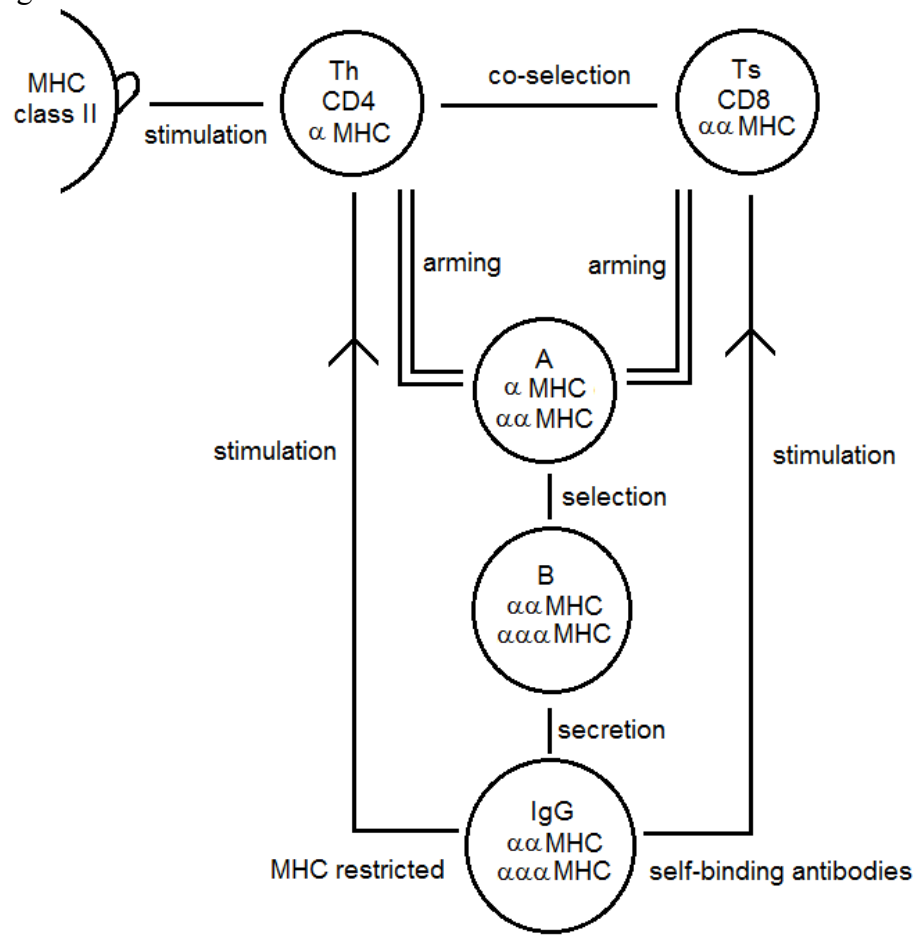

Figure 4